\documentclass[12pt]{article}

\usepackage{sbc-template}
\usepackage{graphicx,url}
\usepackage[utf8]{inputenc}
\usepackage{amsfonts}

\title{A proposal to increase data utility on Global Differential Privacy data based on data use predictions} 

\author{Henry C. Nunes\inst{1}, Marlon P. da Silva\inst{1}, Charles V. Neu\inst{2}, Avelino F. Zorzo\inst{1}}

 \address{Polytechnic School PUCRS, Porto Alegre, Brazil
 \nextinstitute School of Computing Newcastle University, Newcastle upon Tyne, UK
   \email{\{henry.nunes, marlon.pereira\}@edu.pucrs.br}
   \email{charles.neu@newcastle.ac.uk, avelino.zorzo@pucrs.br}
 }

\begin{document} 

\maketitle

\begin{abstract}

This paper presents ongoing research focused on improving the utility of data protected by Global Differential Privacy(DP) in the scenario of summary statistics. Our approach is based on predictions on how an analyst will use statistics released under DP protection, so that a developer can optimise data utility on further usage of the data in the privacy budget allocation. This novel approach can potentially improve the utility of data without compromising privacy constraints. We also propose a metric that can be used by the developer to optimise the budget allocation process.


  
\end{abstract}

\section{Introduction}

Differential Privacy (DP) \cite{dwork:2014} is a new and powerful approach to implement privacy in datasets. It has been used, for example, in government systems \footnote{https://www.census.gov/programs-surveys/decennial-census/decade/2020/planning-management/process/disclosure-avoidance/differential-privacy.html}, and important companies such as Microsoft \footnote{https://azure.microsoft.com/en-us/resources/microsoft-smartnoisedifferential-privacy-machine-learning-case-studies/}and Google \footnote{https://github.com/google/rappor}. However, its application may be difficult, as it frequently requires a tailored solution for a problem, and decreases the data utility due to the anonymization process. Thus, the data utility is a constant problem for DP, being a current research topic the development of approaches that increase data utility after the anonymization process while the desired privacy is preserved.

Our ongoing research proposes a novel approach to improve data utility in DP. We exploit predictions on how the data will be used in the further by the Analyst. These predictions are defined by the developer, the entity that will build the DP solution. The developer is prone to error, missing its predictions can nullify any benefit or even deteriorate the data utility. However, when correctly predicted our approach permits that during the budget allocation process of DP, several queries are privileged. These queries are more influential in how the data will be further used, yielding better utility. Currently, there is no direct way to find the best budget allocation for our approach. In this paper, we propose a metric that the developer can use to help to find better budget allocations by comparing different allocations. However, it is currently a manual process. Our further research includes developing techniques to find such budget allocations automatically. 

Using strategies based on budget allocation to improve utility is not new. There are several approaches for allocation based on the algorithm that will use the data \cite{Fang:2019} \cite{Fan:2019} \cite{Hou:2019}. Other approaches not based on the algorithm that will use the data also exist, such as in the work of Luo \textit{et al.} \cite{luo:2021} inspired on the allocation of resources. However, to the best of our knowledge, there is no other approach that manipulates the budget allocation based on predictions on further use of the data in summary statistics.






\section{Problem Statement}

The \textbf{developer} is an entity that can release several summary statistics to the public using DP. In such situation, there will be a privacy parameter budget $\epsilon$ that is defined externally must be respected by the \textbf{developer}. This budget needs to be divided among all the statistics that will be made available. This \textbf{budget allocation} between different statistics is a pivotal part of our work, and it can be done in different ways \cite{Yan:2020}. 

By the Sequential Composition property sharing the privacy budget keeps the same privacy guarantee \cite{dwork:2014}. To release the data, the \textbf{developer} can build a DP scenario to help in providing the best possible quality of data to the \textbf{Analyst}. 
The \textbf{Analyst} in an entity that intends to consume the data already available to the public. It will use these statistics in equations to generate new insights about the data. The distinction between \textbf{Statistics}, a result from a query in the database with the DP noise added, and the \textbf{Equations} that use multiple statistics from a DP solution to create new insights into the data is important. A third entity is the \textbf{Curator} that works as intended in DP, as an intermediary that will add noise to queries before making them available to the analyst. Here we will treat it as an non-interactive fashion DP solution, as the queries are already defined and are released as \textbf{statistics} to the \textbf{Analyst}.

\begin{figure}[ht]
\centering
\includegraphics[width=1\textwidth]{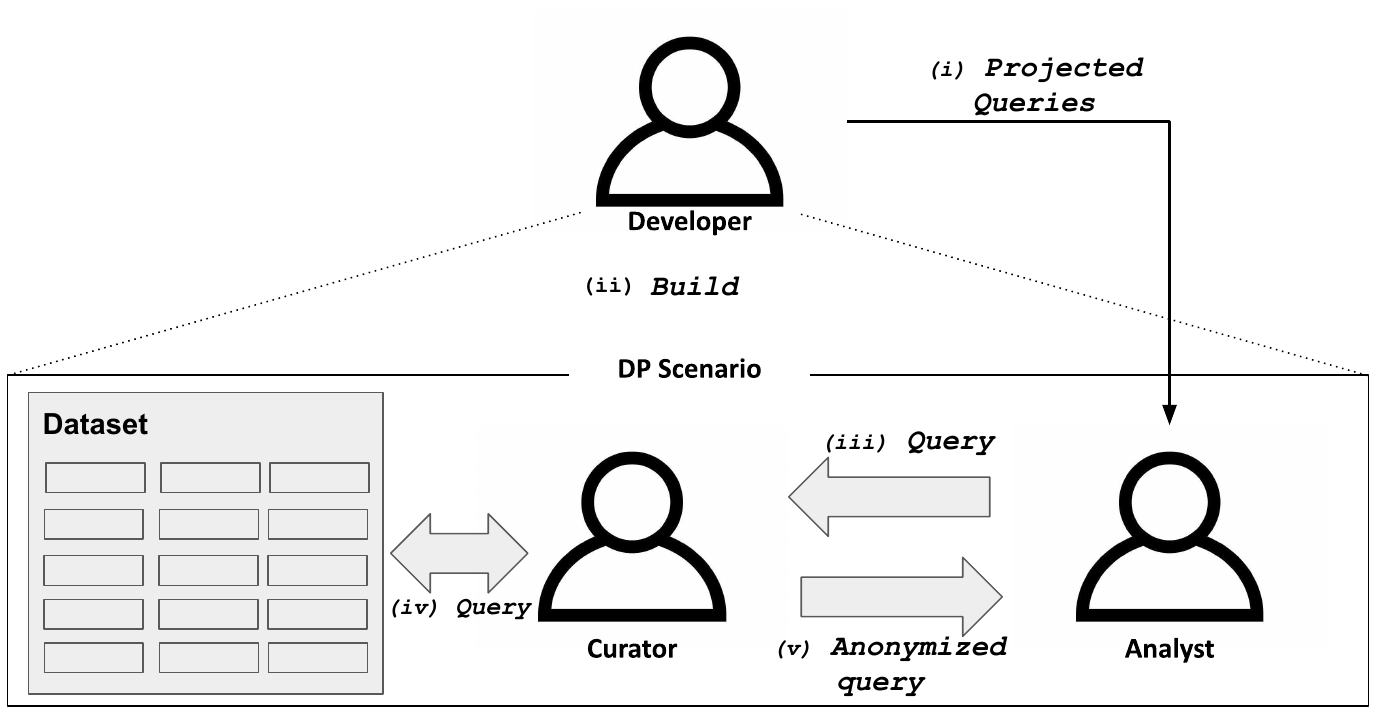}
\caption{Scenario}
\label{fig:scenario}
\end{figure}

Figure \ref{fig:scenario} summarises our scenario. The developer will build the DP scenario. For this, it will project which equations the analyst will create when the DP scenario is built (i). The Developer then creates a \textbf{budget allocation} for all statistics that will benefit these equations by generating the less noise possible. This is a pivotal aspect of our work. 

We will develop this notion in further works, but since each statistic carry its own noise, the interaction between them caused by mathematical operations in equations change the total size of the noise. Depending on the \textbf{budget allocation} this total noise will have a different size. The developer wants to minimize it, which can be done by choosing the optimal \textbf{budget allocation}. In the next section, we present a metric to compare budget allocations that can be used by the developer to compare \textbf{budget allocations} in order to find the optimal solution that minimizes the noise.

Having the optimal budget allocation, it can create the DP scenario (ii). after this, the solution works as normal DP solution (iii) (iv) (v). The Analyst will receive the statistics that the curator did as queries to the database with the added noise. Which the analyst will use in equations. This approach can potentially increase the utility of data without compromising privacy. However, it is highly dependent on the developer's ability to predict the equations that the analyst will use. It improves the utility of the data by prioritizing the predicted equations at the cost of the utility of non-predicted equations. It is also reliant on budget allocation, which is the parameter that the developer can fine-tune. In the next section, we propose a metric that allows comparing different \textbf{budget allocations} in order to optimize the selection.
\section{A Metric to support the process of budget allocation}

In this section we propose a metric that can be used to compare different distributions of the privacy budget $\epsilon$. Initially, an array (1) with a total of $nsta$ statistics that will be released by the developer ranging from $sta_{1}$ to $sta_{nsta}$ is defined.

\begin{equation}
Sta = [sta_{1}, sta_{2}, ...,  sta_{nsta}]
\end{equation}


A second array (2) $Sen$ stores the \textbf{sensitivity} from $sen_{1}$ to $sen_{nsta}$. Each statistic in array $Sen$ has an equivalent member in array $Sta$ at the same position in the equivalent array. For example, the sensitivity for element $sta_{i}$ is $sen_{i}$. 

\begin{equation}
Sen = [sen_{1}, sen_{2}, ...,  sen_{nsta}]
\end{equation}

A third array that we will be used is the budget allocation array (3) $Bud$ with $nsta$ elements. This array represents the privacy budget distribution, for each element in the original array $Sta$ there is an element in $Bud$ in the same position. This represents the privacy budget allocated for that specific statistic. For example, the privacy budget allocated for element $sta_{i}$ is $bud_{i}$.

\begin{equation}
    Bud = [bud_{1}, bud_{2}, ..., bud_{nsta}]
\end{equation}

There are two limitations (4) (5) to the values in this array. The sum of all the elements in $Bud$ must be equal to the privacy budget $\epsilon$.

\begin{equation}
\sum_{i=1}^{nsta} Bud[i] = \epsilon 
\end{equation}

Also, all the elements in the array need to have a value greater than 0. 

\begin{equation}
\forall i (Bud[i] > 0)
\end{equation}

To help organize our data for use in further functions we will consolidate all these three arrays ($Sta$, $Sen$ and $Bud$) in one array of tuples (6) $Tup$. Each tuple will aggregate the statistic, sensitivity, and budget $t = (sta, sen, bud)$. We will also define a function (7) $os$ that retrieves the value $sta$ from a tuple.


\begin{equation}
Tup = [(sta_{1}, sen_{1}, bud_{1}), (sta_{2}, sen_{2}, bud_{2}), ...,  (sta_{nsta}, sen_{nsta}, bud_{nsta})]
\end{equation}

\begin{equation}
os((sta, sen, bud)) = sta
\end{equation}




In the previous section, we described what are \textbf{Equations}. We will now define them as a mathematical function $eq(Tup) \Rightarrow \mathbb{R}$. An equation uses multiple specific statistics to calculate its output value. The equation $eq$ will receive all tuples with their statistics from array $Tup$, although it will use just a few specific ones.

Equations are defined by the developer on a case-by-case base. Thus, it is impossible to define the operation beyond the function signature. However, we will show two examples of equations using lambda function. In our example scenario we have an array $Tup = [t1, t2, t3, t4]$. The developer will define two operations: The first one (8) $eq1$ will receive the statistic value from $t2$, and  $t3$ to sum up their values. In the second one (9) $eq2$, the value of $t1$, $t2$ will be added, and then divided by $t4$. 

\begin{equation}
eq1(Tup) = ((\lambda x\ y. os(x) + os(y)) Tup[2]) Tup[3]
\end{equation}

\begin{equation}
eq2(Tup) = (((\lambda x \ y \ z. \frac{os(x) + os(y)}{os(z)}) Tup[1]) Tup[2]) Tup[3]
\end{equation}

Finally, there will exist a fourth array (10) $Eqs$, that will hold tuples with two values $te = (eq, sen)$. The first value is the function of an equation, defined by the developer. The second one is the sensitivity of that operation if it was a statistic directly retrieved from the database instead of an equation composed of multiple statistics. Further expanding, $eq1$ previously described, is the sum of the statistics from $Tup[2]$, and $Tup[3]$. However, it is possible to get the result of this equation by directly querying the database, which would be the same as retrieve a new statistic. This alternative way of retrieving the value of $eq1$ would have a sensitivity, the value of $sen$ is the sensitivity if instead of using $eq$ we would query the database for a new statistic with the same result as $eq$. For each equation defined by the developer an tuple in $Eqs$ will be created. The size of this array depends of how many equations will be defined by the developer, we define th size as $neq$

\begin{equation}
Eqs = [te_1, te_2, ... te_{neq}]
\end{equation}



To quantify the utility of a single statistic we will create a function $us(t)$ that receive a single tuple from $Tup$. We will not give details about what means utility and how to measure in this work. It will be approached in further work, we make a briefly comment in Section \ref{sec:conclusion}. The important consideration for this work is that it will output a positive real number representing the loss of utility, a higher number means less utility.  


Similar to $us$, we will create another function to quantify the utility of an equation. This function $ue(te)$ receives a tuple from $Eqs$. It outputs a positive real number representing the loss of utility, a higher number means less utility.


Finally, our metric (11) will receive as input the array $Tup$, and $Eqs$. The output is a score that represents the utility for a budget allocation $Bud$, lower means a better utility. The metric function is defined as follows.

\begin{equation}
Metric(Tup, Eqs) = \sum_{i=1}^{nsta} us(Tup[i])  + \sum_{i=1}^{neq} ue(Eqs[i]) 
\end{equation}

Array $Sta$, and $Eqs$ are based on the information that will be disclosed and equations that the developer predicts the analyst will use. This implies that it will not change after being defined. Although, array $Bud$ represents a single instance of all possible divisions of the privacy budget $\epsilon$ to the statistics. The objective of the developer is to find the combination of $Bud$ that yields the lowest metric value. Which means the highest utility. 

One final consideration is why the sensitivity is included in all tuples. The sensitivity can be used to relativize the values of function $us$, and $ue$. A counting query in the database would create a statistic with a sensitivity of one, which would result in a small noise in absolute values when used in a mechanism in DP. While other queries would create a statistic with higher values for sensitivity that could result in bigger noise in absolute values. Both statistics are of equal importance, but the higher noise in absolute value would create a higher return in the function $us$. We plan to use the sensitivity in $us$ and $ue$ to balance all statistics and equations giving them equal importance. This will be developed in further works with the definition of $us$, and $ue$.

\section{Conclusion and Further Work}
\label{sec:conclusion}

In our work, we presented a novel approach to improve the utility of DP scenarios by predicting equations that the analyst will do with the released statistics and benefit them in the budget allocation. To support the approach we presented a new metric system under development that can be used to measure the utility of a specific budget allocation. The approach is designed to be used with summary statistics. 


The potential improvement in utility comes from the budget allocation. Certain allocations increase or decrease the amount of noise that a predicted equation will generate. Using the metric a developer can try to find the allocation that benefits the most taking into consideration all predicted equations. Since it works just on the budget allocation it does not affect the solution privacy.



As a future work, we are evaluating our proposed approach both by theoretical analysis and experimental evaluation. Perhaps the most important is a way of measuring the utility, this measurement would be used in functions $us$, and $uo$ to output utility in a quantitative way. Initially, we are planning to use the amount of expected noise added to a statistic and equation through the DP mechanism as a way to measure utility. The purpose of this measurement could also benefit from a study to evaluate the impact of different budget allocations in simple equations using basic operations, such as sum, multiply, and division, using the proposed measurement. This could be useful to evaluate the measurement and show how different allocations can impact the utility.

The metric now can be used to compare two different budget allocations. An avenue for research is to allow the developer to directly find the optimal budget allocation. Two methods can be explored for this: the first one is using a numerical approach such as gradient descent. The second one is an analytical approach such as a closed mathematical formula. However, further research is needed to evaluate if this approach is feasible.


\bibliographystyle{sbc}
\bibliography{sbc-template}

\end{document}